# Kinetic Analysis of Protein Assembly
# on the Basis of the Center Manifold around the Critical Point

Tatsuaki Tsuruyama*

*Department of Diagnostic Pathology, Kyoto University Graduate School of Medicine

1 Yoshida-Konoe-cho, Sakyo-ku, Kyoto, Kyoto Prefecture 606-8501, Japan

Phone: +81-75-753-4427; Fax: +81-75-753-4493

E-mail: tsuruyam@kuhp.kyoto-u.ac.jp

**Abstract**

Protein assembly plays an important role in the regulation of biological systems. The cytoskeleton assembly activity is provided by the binding cofactors GTP (guanidine triphosphate) or ATP(adenosine triphosphate) to monomeric protein, and is initiated by assembling the monomeric proteins. The binding GTP or ATP is hydrolyzed to GDP (guanidine diphosphate) or ADP (adenosine diphosphate) by the monomeric enzymatic activity. This self-limited assembly is characteristic of the cytoskeleton. To quantitatively evaluate the assembly kinetics, we propose a nonlinear and non-equilibrium kinetic model, with the nonlinearity provided by the fluctuation in monomer concentrations during the diffusion. Numerical simulations suggest that the assembly and disassembly oscillates in a chaos-like manner. We use a kinetic analysis of the center manifold around the critical point to show that minimal increases in ATP/GTP concentrations may lead to some attenuation in the amplification of these fluctuations. The present model and our application of center manifold theory illustrate a unique feature of protein assemblies, and our stability analysis provides an analytical methodology for the biological reaction system.

# 1. Introduction

Protein assembly is essential for cellular activities such as cytoskeleton formation and protein assembly [1, 2]. A transient binding cofactor such as ATP/GTP to monomeric protein controls the assembly of proteins such as actin and tubulin. Various mathematical models have been proposed to explain the kinetic signaling behavior based on a set of kinetic equations [3]. In the current study, we aim to understand the process of protein assembly using the following two novel perspectives, which are formulated on the basis of a nonlinear and non-equilibrium kinetic model. In fact, cytoskeletal protein assembly is a dynamic process between assembly and disassembly [4-8].

According to Fick's law, the continuity of the monomer concentration of $c_i$ ($i = 1,…, n$) including chemical reaction items can be described using diffusion coefficients $D_i$, kinetic coefficients $k_i$, and the concentrations of individual compounds $c_i$. Protein assembly is limited by the slow diffusion rate of monomer proteins, which is a diffusion-limited aggregation process. Therefore, diffusion items and reaction items cannot necessarily be separated and we therefore described kinetic coefficients that depend on the diffusion coefficients as follows[8]:

$$\frac{dc_i}{dt} = k_i D_i c_i + \sum_{j=1}^{n} k_i c_i c_j + \cdots \quad (1)$$

We consider a model system that is far from equilibrium because of a continuous supply of chemicals from the outside. The chemical species are GTP and ATP, cofactors that provide monomers with assembly activity. The aim of this study is to illustrate the behavior of non-equilibrium protein assembly.

# 2. Materials and methods

## 2.1 General formulation of a signaling pathway

The model consists of several steps: (i) the monomer achieves an interactive state by reversibly binding a cofactor (ATP/GTP) that provides the monomer with interaction activity; (ii) the monomer itself has the ability to hydrolyze the cofactor and lose the assembly activity; (iii) the monomer has the ability to exchange the inactive hydrolyzed cofactor (ADP/GDP) with an active non-hydrolyzed one; (v) ATP/GTP are supplied continuously from the outside. The second requirement indicates one of the self-limiting properties of the monomer, causing dynamic instability in the monomer interaction. In the protein interaction kinetics, our analysis of the fluctuation in monomer concentrations was performed using Mathematica 9 (Wolfram Research, Inc., Champaign, IL).

**2.2 Protein interaction kinetics**

The model scheme is shown in Figure 1. There are three types of monomer, ATP/GTP -binding monomer $X$, ADP/GDP--binding monomer $Y$ in the oligomer ($W$), and released ADP/GDP-binding monomer $Z$. $X$ has the higher assembly activity, and $Y$ and $Z$ have the lower assembly activity. We set the oligomer concentration $W$ to be a constant, because *de novo* assembly is considered to be much slower than monomer interaction in the steady state [6, 7, 12-15]. First, $X$ associates with the assembly nucleus $W$ to be $Y$ at the end of $W$.

$$X+W \;\rightarrow\; W+Y \;(m_1; \text{kinetic coefficients}) \qquad (2)$$

In the next step, the intermediate species $Y$ is release to be $Z$:

$$Y \;\rightarrow\; Z \;(m_2) \qquad (3)$$

$Z$ recovers its interaction activity by exchanging the active cofactor ATP/GTP ($P$) for the inactive cofactor ADP/GDP ($P'$), returning to $X$ (see Figure 1):

$$Z+P \;\rightarrow\; X+P' \quad (k_0) \qquad (4)$$

In addition, direct slow conversion is supposed:

$$X \;\rightarrow\; Z \qquad (k_3) \qquad (5)$$

The kinetic equations were set according to the simple reaction cascade described above. We obtain equations for the protein interaction kinetics using the diffusion coefficient:

$$dX/dt = -m_1WX + k_0PZ - k_3X \tag{6'}$$

$$dY/dt = m_1WX - m_2Y \tag{7'}$$

$$dZ/dt = m_2Y - k_0PZ + k_3X \tag{8'}$$

$$X + Y + Z = M \tag{9}$$

A simple consideration of the diffusion-limited step implies that, when the kinetic rate can be described according to Fick's law using the diffusion coefficients $D_X$ and $D_Y$, then[9-11]:

$$m_1 \propto D_X + D_W = k_1 D_1 \tag{10}$$

$$m_2 \propto D_W + D_Y = k_2 D_2 \tag{11}$$

Rewriting (6')-(8') using (10) and (11),

$$dX/dt = -k_1 D_1 WX + k_0 PZ - k_3 X \tag{6}$$

$$dY/dt = k_1 D_1 WX - k_2 D_2 Y \tag{7}$$

$$dZ/dt = k_2 D_2 Y - k_0 PZ + k_3 X \tag{8}$$

To obtain the monomer concentration at the steady state of the reaction system, setting right-hand side of Eqs. (6) – (8) equal to zero and using Eqs. (9) gives:

$$X_e = \frac{D_2 k_2 M p}{D_2 k_3 k_2 + D_2 k_2 p + D_1 D_2 k_1 k_2 W + D_1 k_1 p W} \sim \frac{D_2 k_2 M}{D_2 k_2 + D_1 k_1 W} \tag{12}$$

$$Y_e = \frac{D_1 k_1 M p W}{D_2 k_3 k_2 + D_2 k_2 p + D_1 D_2 k_1 k_2 W + D_1 k_1 p W} \sim \frac{D_1 k_1 M W}{D_2 k_2 + D_1 k_1 W} \tag{13}$$

$$Z_e = \frac{D_2 k_2 M (k_3 + D_1 k_1 W)}{D_2 k_3 k_2 + D_2 k_2 p + D_1 D_2 k_1 k_2 W + D_1 k_1 p W} \sim \frac{D_2 k_2 M (k_3 + D_1 k_1 W)}{D_2 k_2 p + D_1 k_1 p W} \sim 0 \tag{14}$$

Here, the item $D_1 D_2$ and $k_3$ are negligible because the diffusion coefficients are small and the direct conversion rate of $X$ into $Z$ is small.

**2.3 Fluctuation of diffusion coefficient**

We now consider the fluctuation of participant proteins:

$$X = X_e + x,\ Y = Y_e + y,\ Z = Z_e + z \qquad (15)$$

In Eq. (14), the subscript 'e' signifies values at the steady state.

In an actual assembly, monomers associate with other monomers. From Eq. (9), we have $x + y + z = 0$ and therefore the fluctuation $y$ can be written using $-x-z$. Therefore, the fluctuation kinetics are provided by two parameters, $X$ and $Z$.

Given the nonlinearity during diffusion, we assume there is kinetic instability in the monomer interaction, and that the sensitivity of the assembly in response to the environmental change can be evaluated. Indeed, the diffusion coefficient $D$ of one macromolecule in the solution can generally be written as:

$$D = kT/\eta\,(1 - \alpha c) \approx kT\,(1 + \alpha c)/\eta = D_0\,(1 + \alpha c) \qquad (16)$$

where $T$ is the temperature of the solution, $k$ is the Boltzmann constant, $\eta$ is the frictional coefficient of a given macromolecule in solution and $c$ denotes the concentration of the solute, and α is a coefficient, and $D_0$ is the diffusion coefficient [16-20]. The fluctuation of the diffusion coefficient is given by considering the dependency of the coefficients upon the concentration of the monomer [8]:

$$dD_x = \alpha x - \beta z \qquad (17)$$
$$dD_z = \gamma x - \delta z \qquad (18)$$

Here, an increase in $X$ contributes to a decrease in $D_x$ and $D_z$ in the fluctuation term $\alpha x$ ($\alpha > 0$) and $\gamma x$ ($\gamma > 0$), because the higher assembly activity reduces diffusion. In contrast, an increase in $Z$ serves to increase the diffusion coefficients $D_x$ in the fluctuation terms $\beta z$ ($\beta > 0$) and $\delta z$ ($\delta > 0$), because the lower interaction or assembly activity increases diffusion. In fact the assembly activity of $Z$ is lower, the fluctuation item $\delta z$ is negligible. Because $x + y + z = 0$, Eqs (6), (8), (17), and (18) give the fluctuation kinetic equations :

$$dx/dt = (-k_1 D_x W + k_1 \alpha W X_e - k_3)x + (-k_1 \beta W X_e + k_0 P)z + k_1 W \alpha x^2 - k_1 W \beta xz \qquad (19)$$
$$dz/dt = (k_3 - k_2 \gamma Y_e)x - k_0 P z + k_2 \gamma x^2 + k_2 \gamma xz \qquad (20)$$

To simplify the notation, we set:

$$k_1 D_x = D_1, \ k_1 \beta = b, \ k_1 \alpha = a, \ k_2 \gamma = c, \ k_0 P = p, \ k_3 = k \tag{21}$$

and obtain:

$$dx/dt = (-D_1 W + aWX_e - k)x + (-bWX_e + p)z + Wax^2 - Wbxz \tag{22}$$
$$dz/dt = (k - cY_e)x - pz + cx^2 + cxz \tag{23}$$

Around the steady state $(x, z) = (0, 0)$ of Eqs. (22) and (23), the Jacobian matrix of $(dx/dt, dy/dt)$ is given by:

$$L = \begin{bmatrix} -D_1 W + aX_e - k & -bWX_e + p \\ k - cY_e & -p \end{bmatrix} \tag{24}$$

## 3. Results

### 3.1 Calculus simulation of concentration oscillations

The time-course of the monomer concentrations was simulated by substituting appropriate numerical values into Eqs. (22) and (23). A numerical calculation was performed over a sufficiently long period to evaluate the assembly trend. The steady-state concentrations of $X$ and $Z$ are given by Eqs. (12) and (14). The simulation results under the above conditions are shown in Figure 2. When $p$ varies about the value given by

$$\det[L] = (-D_1 W + aX_e - k)(-p) - (k - cY_e)(-bWX_e + p) = 0 \tag{25}$$

Next, we conducted a simulation with values of $M = 0.1$, $W = 1$, $D_1 = 0.28$, $D_2 = 0.012$, $a = 150$, $b = 150$, $k = 0.005$, $c = 0.1$, and $d = 0$. Solving (24) with respect to $p$ with substitution of these values into (24), we find

$$p \approx 0.011 = p_c \tag{26}$$

The small affix c indicates the *critical point* of ATP/GTP concentration. As a result, the fluctuations oscillate between decreasing and increasing as shown in Figure 2. When $p$

< $p_c$, the fluctuation was found to attenuate (Figure 2*d*) and the monomer concentration reached the plateau. While, when $p > p_c$, the fluctuation was found to diverge (Figure 2*f*).

**3.2 Evaluation of model stability using the center manifold around the equilibrium state**

We considered the kinetic stability around the critical point ($p = p_c$) using minimal difference ε. The matrix $L$ is then given by:

$$Lc = \begin{bmatrix} -D_1 W + aX_e - k & -bWX_c + p_c \\ k - cY_e & -p_c \end{bmatrix} \quad (27)$$

Using the eigenvectors of $Lc$, [$l_1$ $l_2$], we perform the following coordinate transformation:

$$\begin{bmatrix} \dot{x} \\ \dot{z} \end{bmatrix} = \begin{bmatrix} l_1 & l_2 \end{bmatrix} \begin{bmatrix} \dot{u} \\ \dot{v} \end{bmatrix} \quad (28)$$

$$\begin{bmatrix} \dot{u} \\ \dot{v} \end{bmatrix} = \begin{bmatrix} l_1 & l_2 \end{bmatrix}^{-1} \begin{bmatrix} \dot{x} \\ \dot{z} \end{bmatrix} \quad (29)$$

In reference to the numerical simulation (Figure 2), when $D_1$, $k$, $p$ are sufficiently small, [$l_1$ $l_2$] ≈ [{-($aWX/k$), 1}, {$aWX/k$, 1}]. We subsequently set:

$$du/dt = f_u(u, v) \quad (30)$$
$$dv/dt = f_v(u, v) \quad (31)$$

The center manifold around the critical point ($p = p_c$) is then given as follows:

$$u = h(\varepsilon, v) = a_1 v^2 + a_2 v\varepsilon + a_3 \varepsilon^2 + a_4 v^3 + a_5 v^2 \varepsilon + a_6 v\varepsilon^2 + a_7 \varepsilon^3 + O(\varepsilon^4) \quad (32)$$

We then analyzed the effect of changes ε in $p$ using the center manifold around the critical point of the system. Subsequently,

$$u = (dv/dt)\partial h(u, \varepsilon)/\partial u + (d\varepsilon/ dt)\partial h(u, \varepsilon)/\partial \varepsilon = (2a_1v + a_2\varepsilon) f_u(u, v) \qquad (33)$$

Using Eqs. (31) and (32), we then obtain:

$$(2a_1v + a_2\varepsilon)f_u(u,v) = a_1 v^2 + a_2 v\varepsilon + a_3\varepsilon^2 + a_4 v^3 + a_5 v^2\varepsilon + a_6 v\varepsilon^2 + a_7\varepsilon^3 + O(\varepsilon^4) \qquad (34)$$

Solving Eq. (34) gives the coefficients of $a_i$ in Eq. (32): $a_3 = a_7 = 0$. Substituting $u$ in Eq. (32) given by $v$ and $\varepsilon$ into $f_v(u,v)$ in Eq. (29), we can obtain the *kinetic stability equation* for fluctuation $v$ using the coefficients $n_i$ ($i = 1, …, 7$) as follows:

$$dv/dt = n_1v^2 + n_2v\varepsilon + n_3\varepsilon^2 + n_4v^3 + n_5v^2\varepsilon + n_6v\varepsilon^2 + n_7\varepsilon^3 + O(\varepsilon^4) \qquad (35)$$

Independent of the numerical values in Eq. (35),

$$n_3, n_6, n_7 = 0 \qquad (36)$$

Then, we have:

$$dv/dt = n_1v^2 + n_2v\varepsilon + n_4v^3 + n_5v^2\varepsilon + O(\varepsilon^4) \qquad (37)$$

By setting left-hand side equivalent to zero,

$$v = 0, \quad \frac{-n_1 - n_5\varepsilon \pm \sqrt{(n_1 + n_5\varepsilon)^2 - 4n_2n_4\varepsilon}}{2n_4} \qquad (38)$$

We have an approximate solution to Eq. (38):

$$v = 0, \approx c\varepsilon, -n_1/n_4 \qquad (39)$$

Here, $c$ is a constant coefficient and two of the oscillatory amplitude of $v$ except $-n_1/n_4$ are nearly equivalent to zero when the concentration of ATP/GTP is equivalent to critical concentration ($\varepsilon \approx 0$). Thereafter the solution $c\varepsilon$ increases as the concentration of ATP/GTP increases. From (32), we have formulation of $u$ using a constant coefficient $c'$,

$$u \approx 0, c'(n_1/n_4)^2 \qquad (40)$$

When $D_1, k, p$ are sufficiently small, substituting $[\mathbf{l_1}\ \mathbf{l_2}] = [\{-aWX/k, 1\}, \{aWX/k, 1\}]$ into (27) gives :

$$x = -(aWX/k)u + v \approx v \qquad (41)$$

As a result, as we described $v$, two of the amplitudes of $x$ are nearly equivalent to zero ($\varepsilon \approx 0$), and another amplitude ($-n_1/n_4$) increases as $\varepsilon$ increases (Figure 2). Thus, stability analysis can predict the behavior of the fluctuation around the critical point of protein assembling system.

## 4. Discussion

We have presented a model for the protein assembly kinetics, and analyzed the stability around the critical point using center manifold theory. The nonlinear kinetic equations include three parameters ($X$, $Y$, and $Z$), but only two are actually independent (Eq.(8)). In the simulations, ATP/GTP or ADP/GDP binding monomers periodically exhibit a chaotic oscillation between assembly and disassembly. This accurately reflects the microtubule kinetics showing instable assembly [4, 5].

To the best of our knowledge, this is one of the first reports on the application of center manifold theory to the analysis of biological reaction systems [8]. The fluctuation of monomer concentrations was subjected to a perturbation expansion using a minimal increase in the supply of ATP/GTP near the concentration at the critical point. This mathematical method can precisely treat nonlinear and multi-parameter systems around the critical point. The fluctuation kinetics are expected to change from convergence to divergence of the concentration fluctuation of the monomer, i.e., stable to instable, around the critical point as shown in Figure 2. Because of this high sensitivity to the concentration of ATP/GTP, the protein assembly is dynamically regulated by minimal changes in the supply of ATP/GTP, which is controlled metabolically. Still, the theoretical basis of protein assembling for quantitative evaluation requires further formulation on the basis of experimental study.

**Acknowledgement**

This work was supported by a Grant-in-Aid for Scientific Research on Innovative Areas, MEXT, Japan. Project name is ''Synergy of fluctuation and Structure''. The funders had no role in study design, data collection and analysis, decision to publish, or preparation of the manuscript.

**Figure Legends**

**Figure 1. Scheme of monomer interaction.** Individual globules or oblongs represent monomers *X, Y, Z,* and oligomer *W*. Kinetic coefficients, $k_0$, $k_1$, $k_2$, and $k_3$ are shown next to the arrows. Outside and inside signify the outside and inside of the cell, respectively. *Y* is located at the end of the oligomer *W*.

**Figure 2. Time-course of the fluctuation in monomer concentrations displays a chaotic oscillation.** Diffusion of active cofactor binding monomer (*X*) and of inactive cofactor binding monomer (*Z*). *p* is (a) 0.000, (b) 0.001, (c) 0.002, (d) 0.004, (e) 0.008, (f) 0.010, (g) 0.010705, and (h) 0.011. The upper graph shows two parametric plots of *X* (red), and *Z* (blue). Lines represent the concentration of *X* and *Z*. The horizontal axis represents time ($0 \leq t \leq 1000$) and the vertical axis represents the concentration of *X* and *Z*. When *p* exceeds 0.01, chaotic oscillations are observed. The Mathematica (version 9, Wolfram Research, Inc., Champaign, IL) code for *p* = 0.01 is as follows: p = 0.01   X = ((D2 M p)/(D2 k + D2 p + D1 D2 W + D1 p W))     Y = ((D1 M p W)/(D2 k + D2 p + D1 D2 W + D1 p W))   Z = ((D2 M (k + D1 W))/(D2 k + D2 p + D1 D2 W + D1 p W))     M = 0.1 W = 1 D1 = 0.28 D2 = 0.0120618556701030931`    a = 150 b = 156 k = 0.005 c = 0.1 d = 0 NDSolve[{Derivative[1][x][t] == -(D1 - a X) x[t] + a x[t]^2 + (p - b X) z[t] - b x[t] z[t] - k x[t], Derivative[1][z][t] == k x[t] + c x[t]^2 + d x[t] z[t] - p z[t],   x[0] == 1.`*^-6, z[0] == 1.`*^-6}, {x, z}, {t, 0, 3300}, MaxSteps -> 50000]    g001 = Plot[{X + x[t]} /. %, {t, 0, 1000}, PlotRange -> All, PlotStyle -> {RGBColor[0, 0, 0]}]     g002 = Plot[{Y - x[t] - z[t]} /. %%, {t, 0, 1000}, PlotRange -> All, PlotStyle -> {RGBColor[1, 0, 0]}]    g003 = Plot[{Z + z[t]} /. %%%, {t, 0,

1000}, PlotRange -> All, PlotStyle -> {RGBColor[0, 0, 1]}, PlotRange -> All]   Show[g001, g002, g003].

**Figure 1**

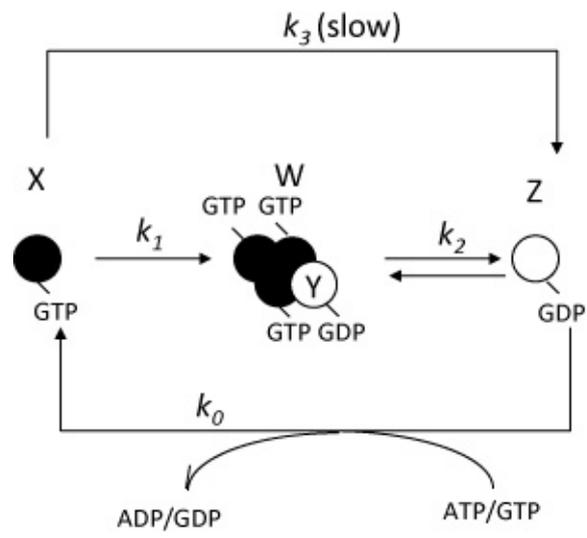

**Figure 2**

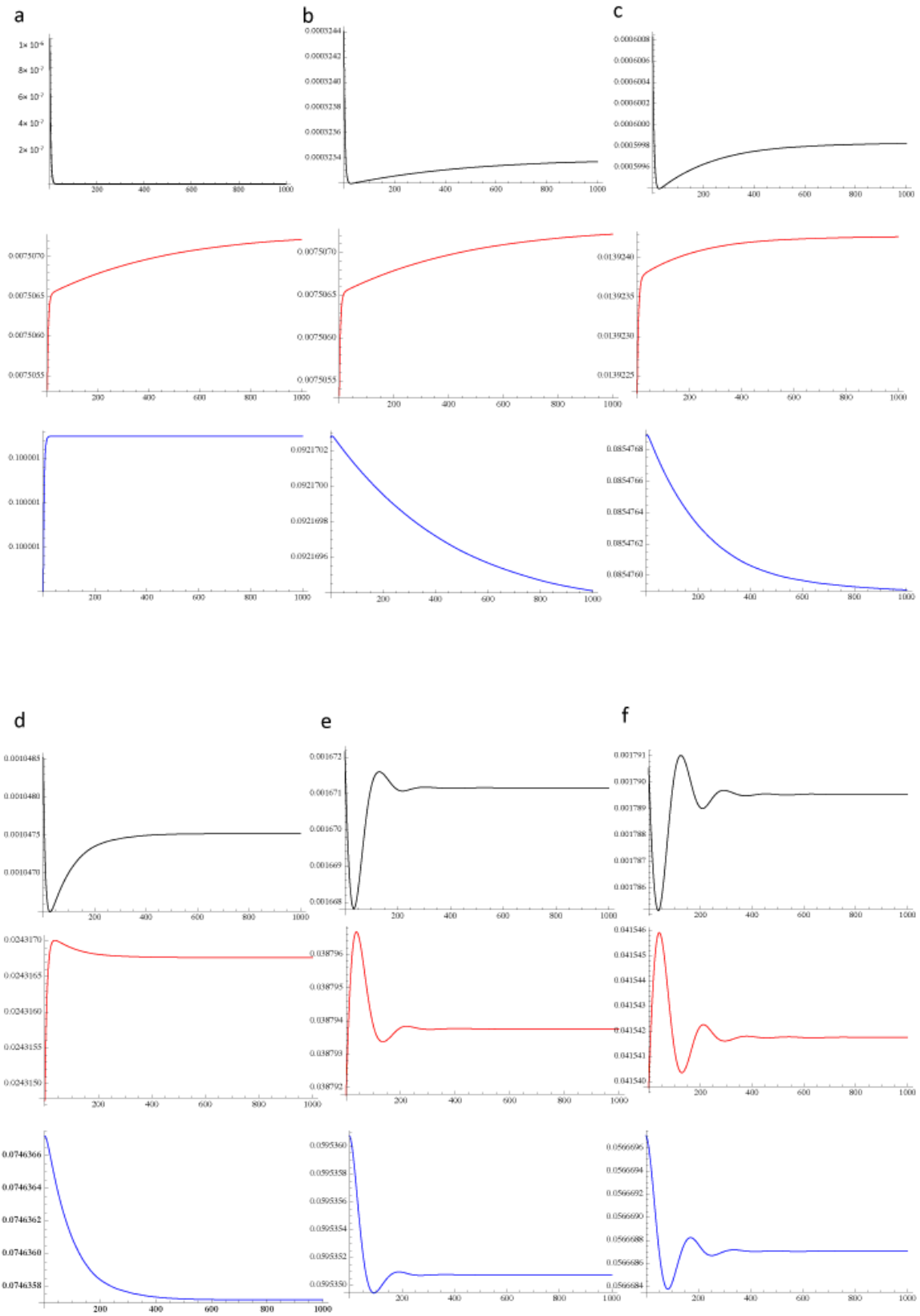

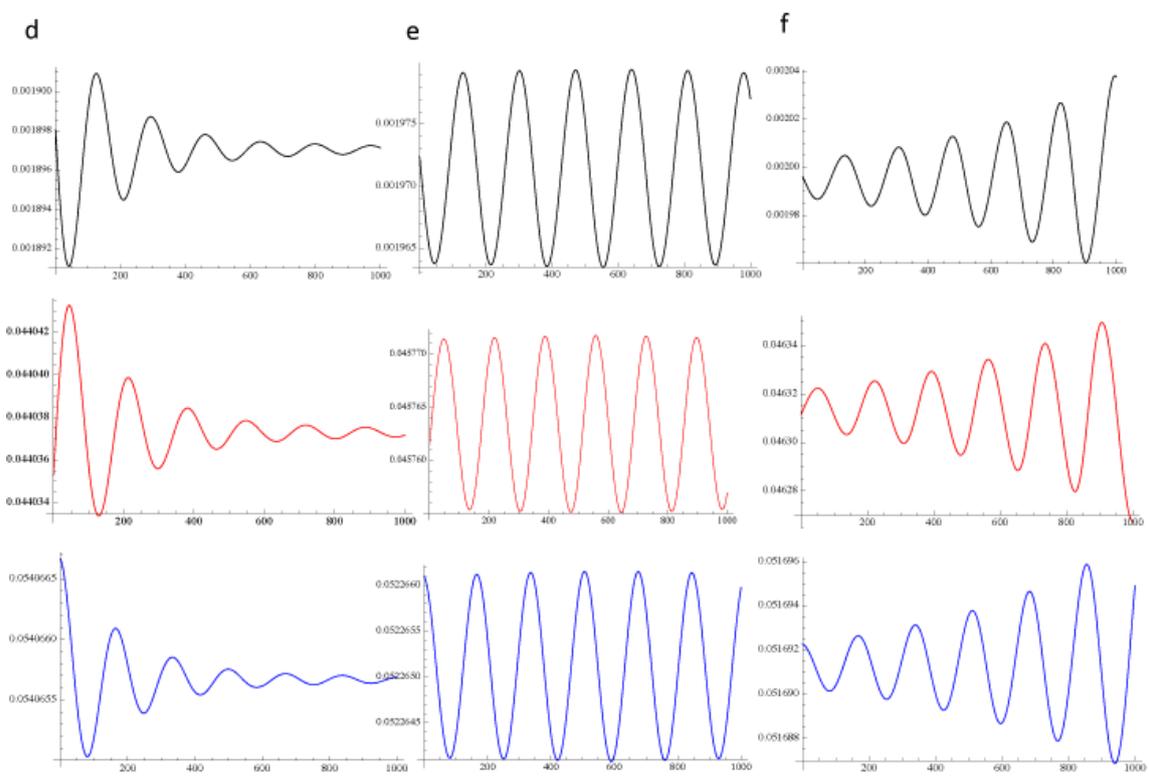